\documentclass{jtitauth}

\begin{document}

\title{Utilization of Reconfigurable Intelligent Surfaces with Context Information: Use Cases}

\author{\L ukasz Ku\l acz}

\markboth{\L ukasz Ku\l acz}
{Utilization of Reconfigurable Intelligent Surfaces with Context Information: Use Cases}

\maketitle

\begin{abstract}
In terms of complex radio environments especially in dense urban areas, a very interesting topic is considered - the utilization of reconfigurable intelligent surfaces. Basically, based on simple controls of the angle of reflection of the signal from the surface, it is possible to achieve different effects in a radio communication system. Maximizing or minimizing the received power at specific locations near the reflecting surface is the most important effect. Thanks to this, it is possible to: receive a signal in a place where it was not possible, detect spectrum occupancy in a place where the sensor could not make a correct detection, or minimize interference in a specific receiver. In this paper, all three concepts are presented, and, using a simple ray tracing simulation, the potential profit in each scenario is shown. In addition, a scenario was analyzed in which several of the aforementioned situations are combined.
\end {abstract}

% \begin{keywords}
% context information, reconfigurable intelligent surfaces, spectrum occupancy detection
% \end{keywords}

\section{Introduction}

In recent times, wireless communication development is based, in many cases, on introducing an ever greater level of intelligence to the network. By the aforementioned intelligence, we mean dynamic spectrum access systems, which often require access to a large amount of contextual information about the environment and network users. This information is necessary to make optimal decisions - based on many parameters (not necessarily directly related to the network itself) and their historical values. Historical data allows, among other things, to capture trends in the behavior of network users and thus the ability to predict upcoming events. The urban environment has always been highly demanding for designers of wireless systems, where the transmitted signal itself is subjected to many propagation phenomena, i.e., reflections or dispersion. The huge number and variety of buildings and objects affecting the propagation of the signal collectively create a highly complex radio channel. Additionally, in an urban environment, there is a~huge number of users in a relatively small area. Combining these two facts is a significant challenge in the operation of wireless systems. 
It is worth noting that, in general, signal reflections are unavoidable. While they hinder the analysis of signal propagation, they are generally a beneficial phenomenon - because the signal can reach the receiver that is not in the direct view of the transmitter. In this context, an exciting approach is the recently popular and considered topic of Reconfigurable Intelligent Surfaces (RIS)~\cite{b1}. At the core of this solution, RIS allows you to influence the process of radio wave reflection on this surface. This means that it is possible to modify the basic principle where the angle of incidence of the wave is equal to the angle of reflection of the wave. This creates great possibilities for creating and controlling the propagation environment. On the one hand, it makes the possibility of directing the radio signal to previously unattainable places (due to the geometry of the objects). On the other hand, the possibility of removing the radio signal in a specific place (minimizing interference) by directing it in a completely different direction~\cite{b2}. Only these two concepts seem to have great potential for use in intelligent networks. They can be used to detect the signal or improve the quality of the received signal (by increasing the received power if it is a desirable signal or reducing the received power if it is an unwanted signal). Signal detection could be possible for a sensor that is somehow hidden from the transmitter in the sense of direct visibility and reflected signals with a sufficiently high power necessary for the operation of signal detection algorithms.
However, the indicated solutions require the development of new algorithms whose task is to control the described RIS~\cite{b3}.

\section{Scenario Description}

In this work, we consider a simple scenario of an elongated room with a wall between the signal transmitter and three receivers, which prevents direct visibility. In the depths of the room, however, there is a surface that will simulate RIS. 

\begin{figure}[!hbp]
\includegraphics[width=0.48\textwidth]{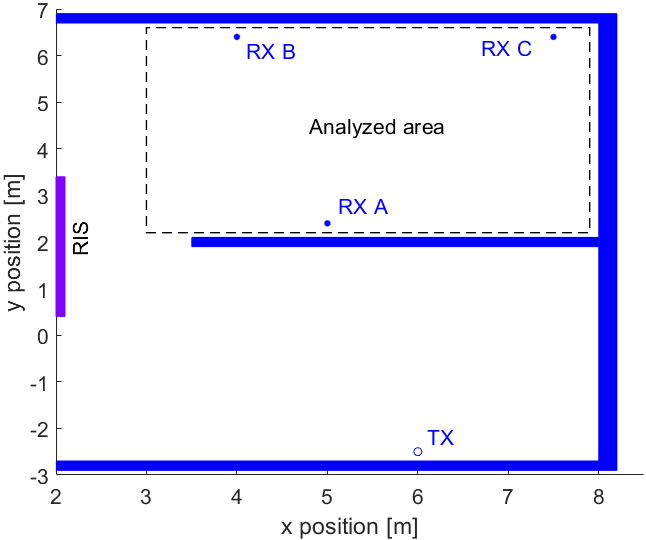}
\caption{Analyzed area}
\label{fig:1}
\end{figure}

All devices are at the same height, and the effects of the ceiling and floor have been ignored - simplifying the analyzed case to two-dimensional space. The projection of the investigated area is presented in Fig.~\ref{fig:1}.
The described room was chosen on purpose as a demonstration of the potential possibilities of using RIS. This decision is dictated by the desire to highlight the concept of using RIS, and the ideas themselves can be used in more complex analyzes, where, however, it would be much more challenging to isolate the impact of RIS. The work described in this paper  compares three simulation scenarios:
\begin{enumerate}
    \item Scenario 1 assumes no use of RIS, so a fragment of the wall was placed instead - this scenario is a reference for the other scenarios.
    \item Scenario 2 assumes a simplified RIS operation that works independently of external factors, i.e., periodically and alternately changes the angle of the reflected signal - which simulates the use of a mirror rotated from left to right and back. This scenario has been proposed to demonstrate the potential profits from using even such a simple control mechanism.    
    \item Finally, Scenario 3 is in which network users have access to information about the current RIS setting and indirectly influence the RIS settings. The last scenario is an example of using context information (the current RIS setting) to achieve the current goal (e.g., minimizing interference). 
 
\end{enumerate}

Three use cases of RIS were analyzed (utilizing all three scenarios) in the paper: improved spectrum occupancy detection by the sensor, improvement in the received signal level for two receivers, and reduction of interference in the receiver. During the simulation, the operation of RIS was simulated as a change in the rotation of the selected surface by an angle from -20 to 20 degrees with a step of 5 degrees.

\section{Simulation Results}

\begin{figure}[!h]
\includegraphics[width=0.48\textwidth]{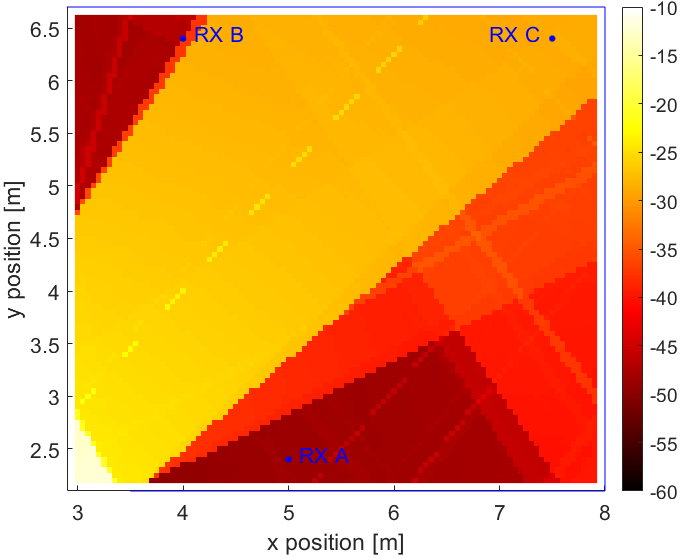}
\caption{Received power without RIS influence}
\label{fig:2}
\end{figure}

Fig.~\ref{fig:2} presents the spatial distribution of the received power in the analyzed area for the given transmitting parameters and uses a propagation model based on ray tracing.

\begin{figure}[!h]
\includegraphics[width=0.48\textwidth]{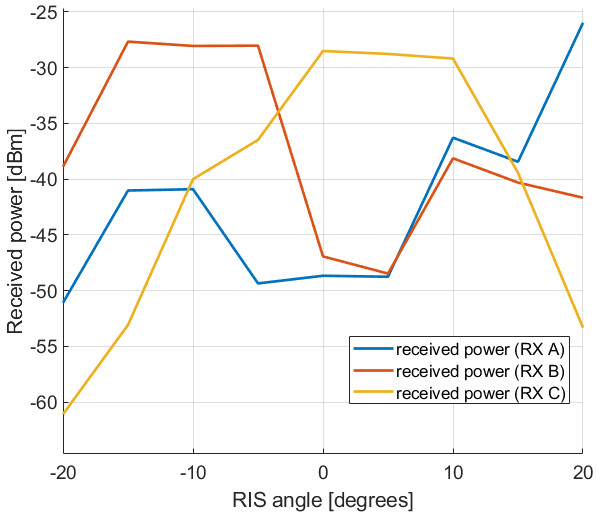}
\caption{Received power in receivers versus RIS setting}
\label{fig:all_angles}
\end{figure}

Preliminary analysis of the indicated distribution shows that part of the area is characterized by less signal attenuation (warmer colors, receiver C). In comparison, the remaining part of the area has greater signal attenuation (darker colors, receivers A and B - remaining "in the shadow" of the signal). The received power in this situation results directly from the perpendicular arrangement of the wall marked as ''RIS'' in Fig.~\ref{fig:1} - however, here plays a role of a wall. It can be seen that receivers A and B would have a problem with receiving and detecting the analyzed signal. However, they would be deprived of excessive interference. The situation is exactly the opposite for receiver C. Nevertheless, it should be remembered that in the opposite situation, when receivers A and B consider the analyzed signal undesirable and receiver C a desirable signal, influencing the current propagation would not be beneficial. Fig.~\ref{fig:all_angles} shows the received power of all three analyzed receivers depending on all considered RIS settings. It shows all possibilities of received power for all receivers, and it will be used depending on the current role of a particular receiver.

\subsection{First Use Case: Spectrum occupancy detection}
\begin{figure}[!hbp]
\includegraphics[width=0.48\textwidth]{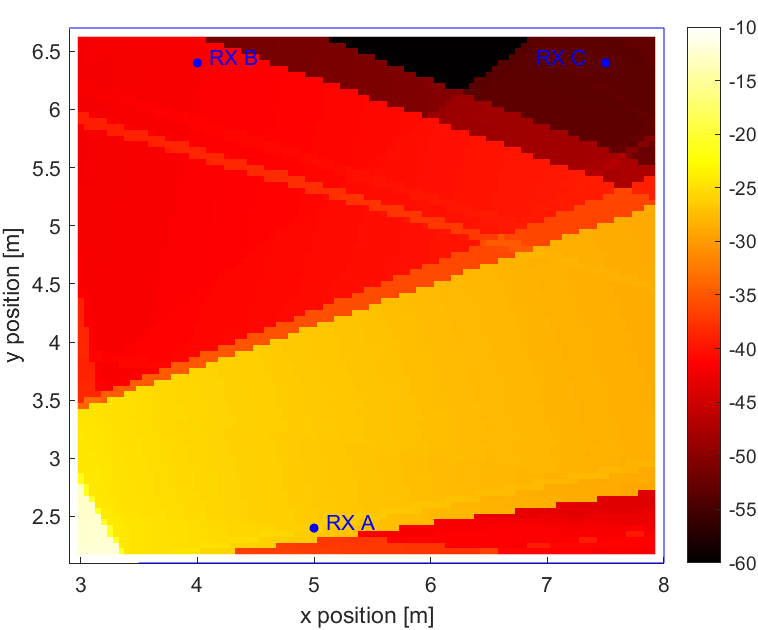}
\caption{Received power map for RIS setting (angle = 20)}
\label{fig:best_sensor}
\end{figure}

For the first use case, i.e., improving the spectrum occupancy detection by receiver B, preliminary simulations showed that in the first scenario, correct detection is challenging; in the second scenario, correct detection is possible for 31.25\% of the time, and in the third scenario, for 79.17\% of time there is a possibility of correct detection. Received power in the second scenario (depending on RIS angle) is presented in Fig.~\ref{fig:all_angles} (see ''RX A''). The last scenario highly depends on the algorithm used (and its purpose). Still, even just information about the current RIS setting allows the sensor to assume - that the occupancy decision is updated only at a particular time. This approach is not ideal, and its effectiveness depends very much on the transmission characteristics (frequency, length of transmission, etc.). The solution, which assumes the ability to control the RIS even in a very simple scenario, enables, for example, periodic "combing" of all available settings and then maintaining one (or several alternating) settings for a certain time. In the considered system with one sensor, it is possible to correctly set the RIS for the entire observation time after a short time, achieving almost 100\% of the possible detection time. Fig.~\ref{fig:best_sensor} presents the received power map with the best RIS setting for receiver A, which is treated like a spectrum detection sensor in this use case.

\subsection{Second Use Case: Received signal level improvement}
\begin{figure}[!h]
\includegraphics[width=0.48\textwidth]{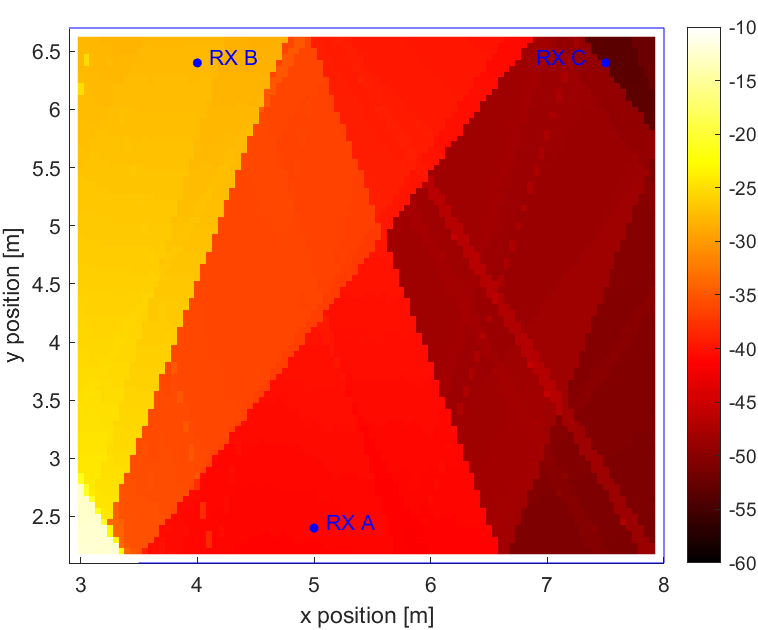}
\caption{Received power map for RIS setting (angle = -15)}
\label{fig:best_b}
\end{figure}

For the second use case, i.e., improving the received signal level for receivers B and C, the preliminary results of the simulation show that in the first scenario, receiver B cannot receive the signal correctly, while receiver C can receive the signal all the time. In the second scenario, 35.29\% of the time, receiver B receives a signal with sufficient power level, while receiver C receives 41.18\% of the time. In this case, we see the profit from the possibility of servicing the receiver, which was in the "shadow" of the signal and could not receive the signal without changing its position. In the third scenario, however, it was assumed that for an initial time, RIS checks all available settings and then, based on reports from receivers, alternately uses the best-reported settings. As a result, receiver B could receive the signal 45.1\% of the time and receiver C 47.06\% of the time. At this point, the possibility of considering traffic steering algorithms to adjust the duration of settings for individual receivers is worth mentioning. For example, on Fig.~\ref{fig:best_b} received power map with the best RIS setting for receiver B was shown. 

\subsection{Third Use Case: Interference reduction}
For the third use case, i.e., interference reduction in receiver C, preliminary simulation results show that in the first scenario, 100\% of the time, the receiver observes significant interference. In the second scenario, the interference exceeds the threshold 35.29\% of the time. It is worth mentioning that the average observed interference level decreases by about 3.715 dB, the median by about 7.972 dB, and the 10th percentile by about 24.762 dB. The 90th percentile remains the same. The knowledge of the RIS setting could at least provide information about the occurrence of potential interference - which can, for example, be used in the transmission planning process. In the third scenario, we again assume a~specified period of checking all RIS settings and then, based on the receiver C report, setting the configuration, causing the minimum level of interference. Thanks to this, it was possible to reduce the interference time above the threshold from 100\% to about 19.61\%. 

\begin{figure}[!h]
\includegraphics[width=0.48\textwidth]{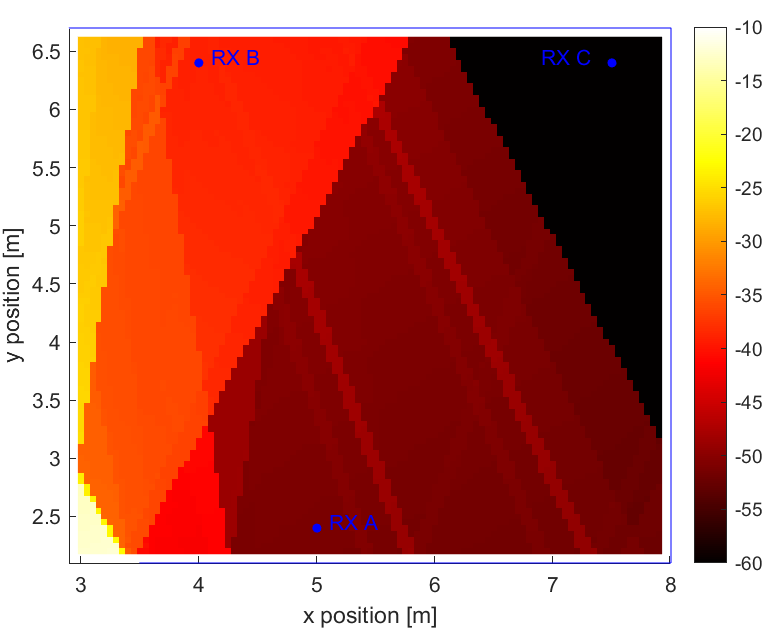}
\caption{Received power map for RIS setting (angle = -20)}
\label{fig:best_c}
\end{figure}

This time, of course, again is hugely dependent on the given configuration and the presence of other devices. Fig.~\ref{fig:best_c} presents the received power map with the best RIS setting for receiver C, which treats the analyzed signal as interference. 
It should be remembered that the simulations carried out assumed constant in-time transmission schema during the whole experiment. In the case of only temporary signal transmissions for all three use cases, this would be an additional challenge for the algorithm that controls the operation of the RIS.

\subsection{All Use Cases combined}
\begin{figure}[!h]
\includegraphics[width=0.48\textwidth]{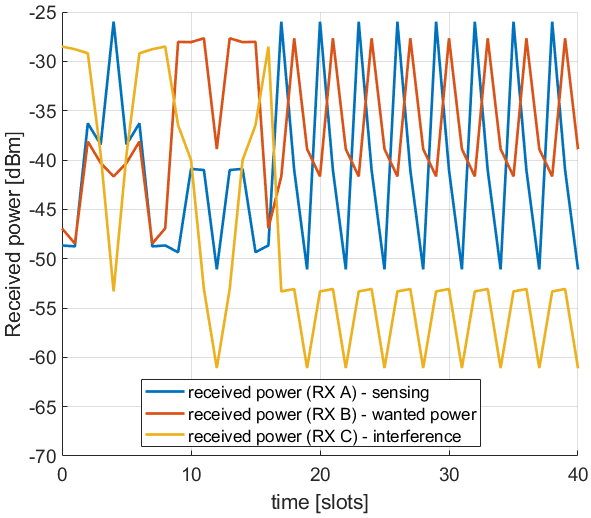}
\caption{Received power in all receivers in time domain}
\label{fig:time}
\end{figure}

In addition, an experiment was carried out combining all three analyzed use cases, i.e., receiver A is a spectrum occupancy detection sensor, receiver B is a receiver of the analyzed signal (desired signal), and receiver C is not a receiver of the analyzed signal (unwanted signal). Recall that in the reference scenario, the objectives of none of the receivers are met. Then, RIS operating according to the following scheme was used: initially, all possible RIS configurations are checked, and then all the best settings reported by the receivers are used alternately. As a result of preliminary simulations, receiver A can successfully detect the signal about 43.4\% of the time; receiver B can successfully receive the desired signal about 45.28\% of the time, and receiver C is free of significant interference about 79.25\% of the time. In addition, the average interference power observed in receiver C decreases by about 8.58 dB. Received power in the time domain for each receiver is shown in Fig.~\ref{fig:time}.

\section{Conclusions}

The simulations presented in the paper show that using RIS, even with the most straightforward control mechanism, can benefit each of the analyzed use cases. However, much better results can be obtained by adding a~simple algorithm to the RIS control mechanism and providing interested receivers with information about the current RIS setting. This gives a very flexible solution that should be carefully considered. In particular, the nature of the environment, the nature of the signal source or sources, potential signal receivers, their positions, and the purpose of use should be considered. Of course, the presented concepts, on the one hand, create the possibility of directing the signal to previously inaccessible places. However, this situation can be both positive and negative because there is a risk of introducing interference in a place where it has not occurred so far. This problem is exacerbated, especially concerning devices unaware of the existence and presence of RIS. Nevertheless, the ideas presented in this paper seem promising regarding potential gains for specific applications.

\section*{Acknowledgements}

The presented work has been funded by the National Science Centre in Poland within the project (no. 2021/43/B/ST7/01365) of the OPUS programme.

\end{document}